\documentclass[%
superscriptaddress,
nofootinbib,
 amsmath,amssymb,
 twocolumn,
]{revtex4}
\usepackage{aas_macros,braket}
\usepackage{bm}
\usepackage{graphicx}
\usepackage{natbib}
\begin{document}
\title{First observational constraints on tensor non-Gaussianity sourced by primordial magnetic fields from cosmic microwave background}
\author{Maresuke Shiraishi}
\email{maresuke.shiraishi@pd.infn.it}
\affiliation{Dipartimento di Fisica e Astronomia ``G. Galilei'', Universit\`a degli Studi di Padova, via Marzolo 8, I-35131, Padova, Italy}
\affiliation{INFN, Sezione di Padova, via Marzolo 8, I-35131, Padova, Italy}
\affiliation{Department of Physics and Astrophysics, Nagoya University, Aichi 464-8602, Japan}
\author{Toyokazu Sekiguchi}
\email{toyokazu.sekiguchi@nagoya-u.jp}
\affiliation{Department of Physics and Astrophysics, Nagoya University,
Aichi 464-8602, Japan}
\date{\today}

\begin{abstract}
Primordial magnetic fields (PMFs) create a large squeezed-type non-Gaussianity in tensor perturbation, which generates non-Gaussian temperature fluctuations in the cosmic microwave background (CMB). We for the first time derive an observational constraint on such a tensor non-Gaussianity from observed CMB maps. Analyzing temperature maps of the WMAP 7-year data, we find that such a tensor non-Gaussianity is consistent with zero. This gives an upper bound on PMF strength smoothed on $1 ~ {\rm Mpc}$ as $B_{1 ~ \rm Mpc} < 3.1 ~{\rm nG}$ at $95\%$ C.L. 
\end{abstract}


\maketitle

%~~~~~~~~~~~~~~~~~~~~~~~~~~~~~~~~~~~~~~~~~~~~~~~~~~~~~~~~~~~~~~~~~~~~~~~~~~~~

%%%%%%%%%%%%%%%%%%%%%%%%%%
\section{Introduction}

Primordial non-Gaussianity is a powerful probe of inflationary models, and various aspects of their property, e.g., amplitude and scale dependence, have been investigated from a diversity of cosmological and astrophysical observables. To date, methods to estimate parameters characterizing non-Gaussianity in primordial perturbations from the cosmic microwave background (CMB) have been extensively investigated by many authors \cite{Komatsu:2001rj, Eriksen:2004df, Smith:2006ud, Yadav:2007rk, Yadav:2007ny, Hikage:2008gy, Matsubara:2010te, Bennett:2012zja, Shiraishi:2013vja}. Some specific types of non-Gaussianity have already been constrained by observed data, e.g., $f_{\rm NL}^{\rm loc} = 2.7 \pm 5.8$, $f_{\rm NL}^{\rm eq} = -42 \pm 75$, and $f_{\rm NL}^{\rm orth} = -25 \pm 39$ (68\%~C.L.) \cite{Ade:2013ydc} (for scale-dependent non-Gaussianities, see Ref.~\cite{Becker:2012je}). 

These bounds have been estimated under an assumption that the primordial non-Gaussianity arises from the scalar perturbations. On the other hand, there exist various models for the early Universe which predict non-Gaussianities associated with not only scalar mode but also vector and tensor modes \cite{Shiraishi:2010kd, Maldacena:2011nz, Soda:2011am, Shiraishi:2011st, Gao:2011vs, Gao:2012ib}. Despite that many attempts have been made so far to constrain primordial non-Gaussianities, those in vector and tensor perturbations predicted by these models have yet to be constrained. Provided predictions from theoretical models and precise data from current CMB observations, we believe that it is timely to investigate constraints on non-Gaussianities in perturbations other than scalar ones. Among various theoretical models, we in this paper focus on the electromagnetic field in the early Universe as a mechanism to generate vector and tensor non-Gaussianities \cite{Shiraishi:2010yk, Kahniashvili:2010us, Shiraishi:2011dh}. 

By cosmological observations of galaxies, cluster of galaxies, and cosmic rays, the existence of large-scale magnetic fields at the present Universe is supported (see, e.g., Refs.~\cite{Wolfe:2008nk, Bernet:2008qp}). There have been a number of studies in which vector fields that exist during inflation are examined as sources for the observed magnetic fields \cite{Grasso:2000wj, Widrow:2002ud, Bamba:2006ga, Martin:2007ue}.  However, due to the problems of backreaction and strong dynamics, it has been in general very difficult to construct a consistent model of magnetogenesis via primordial vector fields \cite{Demozzi:2009fu, Demozzi:2012wh, Suyama:2012wh, Fujita:2012rb, Ringeval:2013hfa}. While these theoretical considerations strongly restrict model building,\footnote{For recent studies of model construction, we refer to, e.g., Refs.~\cite{Ferreira:2013sqa,Kobayashi:2014sga, Caprini:2014mja, Cheng:2014kga}} phenomenological approaches to constrain primordial magnetogenesis are also important. Specifically, through impacts on the CMB anisotropy, properties of primordial magnetic fields (PMFs), which are assumed to be generated from primordial vector fields, can be constrained. For example, observational constraints on the amplitude of PMFs as well as its scale dependence can be obtained by the CMB power spectra alone (for current bounds, see, e.g., Refs.~\cite{Shaw:2010ea, Yadav:2012uz, Paoletti:2012bb, Ade:2013lta}). 

Assuming that the field strength of PMFs has a Gaussian distribution, their energy-momentum tensor creates all types of perturbations, which are highly non-Gaussian due to the quadratic dependence on the field strength \cite{Brown:2005kr, Seshadri:2009sy, Caprini:2009vk,Cai:2010uw, Shiraishi:2010yk, Kahniashvili:2010us, Trivedi:2010gi, Trivedi:2011vt, Shiraishi:2011dh, Shiraishi:2011fi, Shiraishi:2012rm, Trivedi:2013wqa, Shiraishi:2013vha}. This leads to non-Gaussian CMB anisotropies and suggests that higher order correlation functions or polyspectra of the CMB anisotropy beyond the power spectrum should also be informative in probing PMFs. On the basis of this concept, this paper newly explores an observational constraint on the PMF strength by evaluating the magnitude of non-Gaussianity in the CMB temperature anisotropies.

In the case of PMF, the tensor non-Gaussianity, which becomes prominent in the squeezed limit, dominates over the scalar one \cite{Shiraishi:2012rm}, and hence, non-Gaussian temperature fluctuations mainly have information of the tensor mode. Since CMB tensor-mode fluctuations generated from PMFs have unique features and are distinct from CMB signals from ordinary scalar perturbations in inflationary Universe, a nontrivial constraint is expected to be obtained. In this sense, this work corresponds to a first attempt to constrain a tensor non-Gaussianity from CMB data. This is also another motivation of this paper. 

This paper is organized as follows. In the next section, we summarize the tensor non-Gaussianity originating from PMFs. In Sec.~\ref{sec:bound}, after performing some validation tests of our bispectrum estimator and data treatments, we put limits on the magnetic tensor bispectrum from the observed temperature maps of the WMAP 7-year result \cite{Komatsu:2010fb, Larson:2010gs, Jarosik:2010iu, Gold:2010fm}, which we translate into constraints on the amplitude of PMFs. The final section is devoted to summary and discussion.

%%%%%%%%%%%%%%%%%%%%%%%%%% 
\section{Tensor non-Gaussianity generated from PMFs}\label{sec:theory}

First, we briefly summarize the mechanism of PMFs to generate CMB temperature fluctuations and its signatures in observed CMB bispectrum. After PMFs are produced and stretched beyond the horizon during inflation, the anisotropic stress of PMFs contributes to the source term in the Einstein equation and supports the growth of curvature and tensor perturbations even on superhorizon scales until neutrino decoupling. However, subsequent to neutrino decoupling, finite anisotropic stress fluctuations in neutrinos cancel out the magnetic anisotropic stress fluctuations, and therefore, the enhancement of metric perturbations ceases. The resultant curvature and tensor perturbations produce CMB anisotropies, which are called passive-mode fluctuations \cite{Shaw:2009nf}. 

Let us denotes the initial perturbations of the transverse-traceless (TT) part of the metric as
$\delta g_{ij}^{\rm TT} / a^2 = \int \frac{d^3 {\bf k}}{(2\pi)^3} \sum_{\lambda = \pm 2} h_{\bf k}^{(\lambda)}e_{ij}^{(\lambda)}(\hat{\bf k}) e^{i {\bf k} \cdot {\bf x}}$, 
where $\lambda$ denotes the helicity and $e_{ij}^{(\lambda)} (\hat{\bf k})$ is the basis of TT tensors
obeying $e_{ij}^{(\lambda)}(\hat{\bf k}) e_{ij}^{(\lambda')}(\hat{\bf k}) = 2 \delta_{\lambda, -\lambda'}$ and 
$e_{ij}^{(\lambda) *}(\hat{\bf k}) = e_{ij}^{(-\lambda)}(\hat{\bf k}) = e_{ij}^{(\lambda)}(- \hat{\bf k})$ \cite{Shiraishi:2010kd}. 
The initial condition of CMB fluctuations is determined by $h^{(\lambda)}_{\bf k}$, which are estimated as \cite{Shaw:2009nf}
\begin{eqnarray}
h_{\bf k}^{(\lambda)} &\approx& - 1.8 
\frac{\ln(T_B / T_\nu )}{4\pi \rho_{\gamma,0}} 
e_{ij}^{(-\lambda)}(\hat{\bf k})  \nonumber \\ 
&&\times \int \frac{d^3 {\bf k'}}{(2 \pi)^3} B_i({\bf k'}) B_j({\bf k} - {\bf k'}) ~, \label{eq:GW}
\end{eqnarray}
where $T_B$ and $T_\nu$ denote the energy scales at PMF creation and neutrino decoupling, respectively, and $\rho_{\gamma,0}$ is the present photon energy density. Supposing that PMFs $B_i$ are quantum-mechanically created and the probability distribution of their field strength obeys pure Gaussian statistics as is the case in majority of models, $h_{\bf k}^{(\lambda)}$, which is proportional to the PMF anisotropic stress fluctuations, becomes highly non-Gaussian fields obeying the chi-square distribution due to the quadratic dependence on the Gaussian PMFs. Owing to the local form of Eq.~\eqref{eq:GW}, the bispectrum of gravitational waves [$\Braket{\prod_{n=1}^3 h_{{\bf k}_n}^{(\lambda_n)} }$ or $\Braket{\prod_{n=1}^3 h_{\ell_n m_n}^{(\lambda_n)}(k_n) }$ in Eq.~\eqref{eq:CMB_bis}] is amplified in the squeezed limit ($k_1 \approx k_2 \gg k_3$ or $\ell_1 \approx \ell_2 \gg \ell_3$) if the PMF power spectrum is nearly scale invariant \cite{Shiraishi:2011dh, Shiraishi:2012rm}. 

The CMB temperature anisotropies for given direction $\hat{\bf n}$ are quantified via the spherical harmonics expansion as $\frac{\Delta T(\hat{\bf n})}{T} = \sum_{\ell m} a_{\ell m} Y_{\ell m}(\hat{\bf n})$. Using a harmonic-space representation, $h_{\bf k}^{(\lambda)} = \sum_{\ell m} h_{\ell m}^{(\lambda)}(k) {}_{-\lambda}Y_{\ell m}(\hat{\bf k})$, the CMB bispectrum is formed as \cite{Shiraishi:2010sm, Shiraishi:2010kd}
\begin{eqnarray}
\Braket{\prod_{n=1}^3 a_{\ell_n m_n}} 
&=& \left[ \prod_{n=1}^3 (-i)^{\ell_n} 
\int \frac{k_n^2 dk_n}{2 \pi^2} {\cal T}_{\ell_n}(k_n) 
\right] \nonumber \\ 
&& \times \sum_{\lambda_1, \lambda_2, \lambda_3 = \pm 2} 
\Braket{\prod_{n=1}^3 h_{\ell_n m_n}^{(\lambda_n)}(k_n) } \nonumber \\
&=& 
\left(
  \begin{array}{ccc}
  \ell_1 & \ell_2 & \ell_3 \\
  m_1 & m_2 & m_3
  \end{array}
 \right) B_{\ell_1 \ell_2 \ell_3}~, \label{eq:CMB_bis}
\end{eqnarray}
where ${\cal T}_{\ell}(k)$ is the temperature transfer function of the tensor mode involving the amplification for $\ell \lesssim 100$ by the tensor-mode Integrated Sachs-Wolfe (ISW) effect \cite{Pritchard:2004qp}. The transfer function determines the shapes of the CMB bispectrum; thus, the tensor-mode magnetic $B_{\ell_1 \ell_2 \ell_3}$ under examination is less correlated with the usual scalar local-type one, even if their primordial Fourier-space bispectra resemble each other \cite{Shiraishi:2012rm, Shiraishi:2013vha}. According to Ref.~\cite{Shiraishi:2012rm}, under the presence of PMFs, the tensor mode dominates over total signal of the CMB bispectrum at the WMAP angular resolution ($\ell \lesssim 500$), and the contributions of scalar and vector modes are negligible. Therefore, in our bispectrum estimation, we take into account the signals coming from the tensor non-Gaussianity \eqref{eq:GW} alone. Note that $B_{\ell_1 \ell_2 \ell_3}$ is proportional to the magnetic field strength to the sixth power. 

In what follows, we obey the conventional parametrization for the power spectrum of PMFs as 
\begin{eqnarray}
\Braket{B^i({\bf k}) B_j({\bf k'})} 
&=& (2\pi)^3 \frac{P_B(k)}{2} P^i_{~j}(\hat{\bf k}) \delta^{(3)} ({\bf k} + {\bf k'}) ~, \\ 
%------
P_B(k) &=& \frac{(2\pi)^{n_B + 5} B_{1 ~\rm Mpc}^2}{ \Gamma(\frac{n_B + 3}{2}) (\frac{2\pi}{1~ \rm Mpc})^{n_B + 3}} k^{n_B},
\end{eqnarray}
where $P^i_{~j}(\hat{\bf k}) \equiv \delta^i_{~j} - \hat{k}^i \hat{k}_j$, $n_B$ and $B_{1 ~ \rm Mpc}$ are the divergence-free projection tensor, the PMF spectral index, and the PMF strength smoothed on $1~ \rm Mpc$ scale, respectively. In the next section, we constrain the amplitude of bispectrum given by 
\begin{eqnarray}
A \equiv \left(\frac{B_{1 ~\rm Mpc}}{3 ~{\rm nG}}\right)^6 \propto B_{\ell_1 \ell_2 \ell_3} ~,
\end{eqnarray}
under an assumption of the generation of PMFs at the GUT scale ($T_B / T_\nu = 10^{17}$) and nearly scale-invariant shapes of the PMF power spectrum ($n_B = -2.9$). Note that theoretically $A$ should take a positive value. 

In the {\it Planck} bispectrum analysis \cite{Ade:2013ydc}, a non-Gaussianity parameter of the Legendre-polynomial bispectra (the so-called $c_2$), parametrizing the size of magnetic bispectrum of the passive scalar mode with $n_B = -2.9$ and $T_B / T_\nu = 10^{17}$, namely $c_2 \approx -2A$, was constrained, and the corresponding limit is $A = -1.9 \pm 13.9$ (68\% C.L.) or $B_{1 ~ \rm Mpc} < 5.2 ~{\rm nG}$ (95\% C.L.) \cite{Shiraishi:2013vja}. In the next section, we obtain more stringent constraints even from the WMAP data since the most dominant contribution comes from the tensor-mode bispectrum under examination, which was not included in the previous analysis \cite{Ade:2013ydc}. We note that there are other bounds on PMFs from the CMB power spectra, where the so-called magnetic compensated mode \cite{Paoletti:2008ck, Shaw:2009nf} is included and/or $n_B$ is treated as a free parameter: Planck gives $B_{1 ~ \rm Mpc} < 3.4 ~{\rm nG}$ and $n_B < 0$ \cite{Ade:2013lta}, and SPT gives $B_{\rm 1~Mpc} < 3.5~{\rm nG}$ \cite{Paoletti:2012bb}.

%%%%%%%%%%%%%%%%%%%%%%%%%% 

\section{Observational limits}\label{sec:bound}

Given a theoretical template of $B_{\ell_1 \ell_2 \ell_3}$ for a specific theoretical model, we can in general construct an optimal estimator of the amplitude of the bispectrum of primordial perturbations \cite{Komatsu:2003iq, Yadav:2007ny}. 
An optimal cubic estimator for the amplitude of bispectrum can be approximated as
\begin{eqnarray}
\hat{A} &=& \frac{1}{6{\mathcal{N}}}
\sum_{\ell_i m_i} 
\left(
\begin{array}{ccc}
\ell_1 & \ell_2 & \ell_3 \\
m_1 & m_2 & m_3
\end{array}
\right)
B_{\ell_1 \ell_2 \ell_3}^{A=1} \nonumber \\  
&&\times \left[ \prod_{n=1}^3 \frac{a_{\ell_n m_n}}{C_{\ell_n}} 
 - 3  \frac{\Braket{a_{\ell_1 m_1} a_{\ell_2 m_2}}_{\rm MC}}{C_{\ell_1} C_{\ell_2}} \frac{a_{\ell_3 m_3}}{C_{\ell_3}} \right], \label{eq:estimator}
\end{eqnarray}
where $B_{\ell_1 \ell_2 \ell_3}^{A=1}$ is the template bispectrum for the PMF model being normalized with $A = 1$, the bracket denotes the ensemble average of (Gaussian) Monte-Carlo realizations, and $\mathcal N$ is the normalization factor equal to the Fisher matrix: 
\begin{eqnarray}
{\cal N} &\equiv& \sum_{\ell_1 \ell_2 \ell_3} 
\frac{(B_{\ell_1 \ell_2 \ell_3}^{A=1})^2 }{6C_{\ell_1} C_{\ell_2} C_{\ell_3}} ~.
\end{eqnarray}
This estimator form is derived under the so-called diagonal covariance approximation where numerically unfeasible computations of inverse of the covariance matrix are avoided by a simple replacement $(C^{-1} a)_\ell \to a_\ell/C_\ell$. Practically, the bispectrum estimations based on this approximate form and the simple recursive inpainting technique for regions covered by mask retain optimality (with error bars that agree with the optimal ones derived from the Fisher matrix within 5\%), and hence, it has been adopted in the {\it Planck} analysis \cite{Ade:2013ydc}. Note that this form automatically involves relevant experimental features, i.e., beam, partial sky mask, and anisotropic noise, as $B_{\ell_1 \ell_2 \ell_3} = b_{\ell_1} b_{\ell_2} b_{\ell_3} B_{\ell_1 \ell_2 \ell_3}^{\rm theory}$ and $C_\ell = b_\ell^2 C_\ell^{\rm theory} + N_\ell$, with $b_\ell$ and $N_\ell$ denoting beam transfer function and noise spectrum. 

The form \eqref{eq:estimator} indicates that to obtain $\hat A$ from a single realization, a direct implementation requires ${\cal O}(\ell_{\rm max}^5)$ arithmetics (as is the case in non-Gaussian map creation mentioned in Sec.~\ref{sec:test}), where $\ell_{\rm max}$ is the maximum multipole. For $\ell_{\rm max} \sim 1000$, required computational time is enormous. In the literature of the so-called KSW approach \cite{Komatsu:2003fd, Senatore:2009gt, Creminelli:2005hu, Yadav:2007ny}, a factorized estimator form has been found for the case of the standard scalar non-Gaussianity where the angle dependence is removed, since the dependence on $(\ell_1, m_1)$, $(\ell_2, m_2)$, and $(\ell_3, m_3)$ can be separated from one another. On the other hand, in the tensor case, due to complicated spin dependence, different multipoles are tangled with one another and there is no way to reduce numerical operations in the same manner as the KSW approach.\footnote{The so-called separable modal estimator \cite{Fergusson:2009nv, Fergusson:2010dm, Liguori:2010hx, Fergusson:2011sa, Shiraishi:2014roa, Fergusson:2014gea, Shiraishi:2014ila, Liguori:2014} is applicable to general nonfactorizable bispectrum templates like the PMF case.} In this paper, we straightforwardly perform ${\cal O}(\ell_{\rm max}^5)$ arithmetics in estimator computations; then to prevents us from taking too much computational time, let us stop summations at $\ell_{\rm max} = 100$. In our PMF case, the signal-to-noise ratio is almost saturated at $\ell \simeq 100$ \cite{Shiraishi:2012rm}; thus, we believe that by choosing such small $\ell_{\rm max}$, constraints on $A$ do not change so much in comparison with analyses at the WMAP resolution $\ell_{\rm max} \simeq 500$.

A summary of our analysis and treatment of the data set are as follows. In Sec.~\ref{sec:WMAP}, we place observational limits on the magnetic bispectrum using the coadded temperature maps from the WMAP 7-year observation at V and W bands \cite{Jarosik:2010iu,Gold:2010fm}.\footnote{http://lambda.gsfc.nasa.gov} We then compare the constraints from both (not foreground-cleaned) raw and foreground-cleaned data. Prior to it, in Sec.~\ref{sec:test}, we check the validity of our estimator by using simulated non-Gaussian maps originating from known magnetic bispectrum. In these works, for error estimations and linear term computations, we use 500 simulated Gaussian maps. Taking into account experimental uncertainties, in these maps, we include an anisotropic noise component. Furthermore, to reduce effects of residual foregrounds, we apply the KQ75y7 mask recommended by the WMAP team \cite{Gold:2010fm}, whose sky coverage is $f_{\rm sky} = 0.706$. After removing monopole and dipole components, the masked regions are inpainted by means of the recursive inpainting procedure adopted in the {\it Planck} analysis~\cite{Ade:2013ydc}. Our pixel-space computations are based on a resolution $N_{\rm side}=512$ in the HEALPix pixelization scheme \cite{Gorski:2004by}.\footnote{http://healpix.jpl.nasa.gov} The CMB signal power spectrum $C_\ell$ is computed using the {\tt CAMB} code \cite{Lewis:1999bs}, assuming a concordance flat power-law $\Lambda$CDM model with the mean cosmological parameters from the WMAP 7-year data alone \cite{Komatsu:2010fb}. Our beam transfer function and anisotropic noise component are generated by coadding the data in the V and W band channels by means of the WMAP-team method \cite{Komatsu:2008hk}.

%*********************
\subsection{Validation tests using simulated maps} \label{sec:test}

Before moving to the actual data analysis, we check the validity of our bispectrum estimations mentioned above using simulated non-Gaussian maps with known PMF bispectrum. More specifically, we generate 50 realizations of non-Gaussian CMB temperature maps assuming $A = 3.50$, which corresponds to $\sim 3 \sigma$ significance and compare the estimator $\hat A$ of Eq.~\eqref{eq:estimator} 
from these realizations with the input $A$.

\begin{figure}[t]
\begin{tabular}{c}
    \begin{minipage}{1.0\hsize}
  \begin{center}
    \includegraphics[width = 1\textwidth]{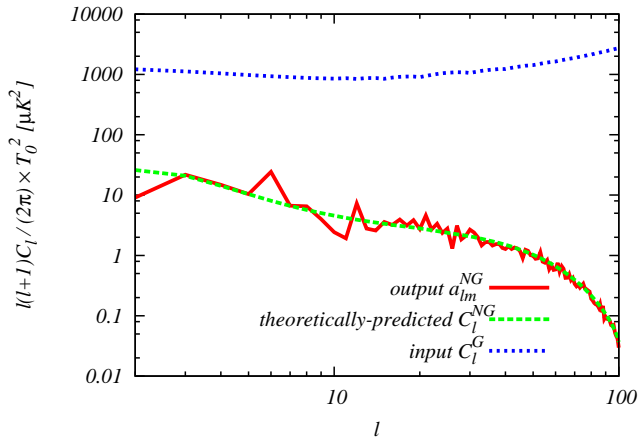}
  \end{center}
\end{minipage}
\end{tabular}
  \caption{Angular power spectrum of a single non-Gaussian realization $a_{\ell m}^{\rm NG}$ with $A = 3.50$ (red solid) generated from the input Gaussian power spectrum $C_\ell^{\rm G}$ (blue dotted). The consistency with the theoretically predicted power spectrum of $a_{\ell m}^{\rm NG}$ (green dashed) is confirmed. It is visually apparent that $a_{\ell m}^{\rm NG}$ decays rapidly at around $\ell = 100$ due to the end of the ISW enhancement in the tensor bispectrum.}
  \label{fig:Cl_sim}
\end{figure}

According to Refs.~\cite{Fergusson:2009nv, Hanson:2009kg, Curto:2011zt}, given a power spectrum $C_\ell$ and bispectrum $B_{\ell_1 \ell_2 \ell_3}$, a random realization of CMB temperature anisotropy $a_{\ell m}$ can be approximately given as
\begin{eqnarray}
a_{\ell m} &\equiv& a_{\ell m}^{\rm G} + a_{\ell m}^{\rm NG} ~, \\
%--- 
a_{\ell_1 m_1}^{\rm NG} &=& \frac{1}{6} 
\left[ \prod_{n=2}^3 \sum_{\ell_n m_n} 
\frac{a_{\ell_n m_n}^{{\rm G}*}}{C_{\ell_n}} \right] \nonumber \\
&&\times 
\left(
  \begin{array}{ccc}
  \ell_1 & \ell_2 & \ell_3 \\
  m_1 & m_2 & m_3
  \end{array}
 \right)
B_{\ell_1 \ell_2 \ell_3} 
~. 
\label{eq:almNG_ex}
\end{eqnarray}
Here, $a_{\ell m}^{\rm G}$ is the Gaussian part of a realization whose variance is given by $\Braket{\prod_{n=1}^2 a_{\ell_n m_n}^{\rm G}} = C_{\ell_1} (-1)^{m_1} \delta_{\ell_1, \ell_2} \delta_{m_1, -m_2}$, and $a_{\ell m}^{\rm NG}$ denotes the non-Gaussian part of the realization. In the same manner as the estimator computation, we stop summations at $\ell_{\rm max} = 100$. This truncation is reasonable  since, for $\ell \gtrsim 100$, the tensor bispectrum is highly damped and does not contribute to $a_{\ell m}^{\rm NG}$ as shown in Fig.~\ref{fig:Cl_sim}. This also enables us to generate many non-Gaussian maps, despite the need for ${\cal O}(\ell_{\rm max}^5)$ operations. 

Mean values of $A$ computed from 50 non-Gaussian maps with $A = 3.50$ and $1\sigma$ errors estimated from 500 Gaussian maps are summarized in Table~\ref{tab:test}. In these estimations, we assume two types of surveys: a full-sky noiseless ``ideal'' survey and a ``WMAP-like'' survey involving all experimental features of WMAP discussed above. It is verified from Table~\ref{tab:test} that as expected, our estimator recovers the input value $A = 3.50$ within error bars both in the ideal and WMAP-like surveys. Moreover, the resultant error bars are well consistent with the Fisher matrix values: $\delta A = 1/\sqrt{f_{\rm sky} {\cal N}} = 1.17$ (ideal) and $1.39$ (WMAP). These results show that our estimator retains optimality and support the validity of our computations.   

\begin{table}[t]
\begin{center}
  \begin{tabular}{|c||c|c|c|c|c|c|c|c|} \hline
    & Ideal: $f_{\rm sky} = 1$ & WMAP: $f_{\rm sky} = 0.706$ \\ \hline
    Average & 3.54 & 3.86 \\ \hline
    $1\sigma$ error & 1.13 & 1.36 \\  \hline
  \end{tabular}
\end{center}
\caption{
Mean values of $A$ obtained from 50 simulated non-Gaussian maps with the input value $A = 3.50$, together with $1\sigma$ errors estimated from 500 Gaussian maps. We here compute the estimator assuming full-sky noiseless ideal and WMAP-like surveys. It is found that the mean values and the error bars reach the input value $3.50$ and the Fisher matrix values $1.17$ (ideal) and $1.39$ (WMAP), respectively. 
} 
\label{tab:test}
\end{table}

%*********************

\subsection{WMAP results} \label{sec:WMAP}

Here we present our constraints from the WMAP7 (raw and foreground-cleaned) data, including experimental features (beam, noise and mask) and inpainted as mentioned above. Prior to estimating the amplitude of the magnetic bispectrum, we estimated $f^{\rm loc}_{\rm NL}$ with $\ell_{\rm max} = 100$ from the foreground-cleaned data and found the $1\sigma$ bound: $(-1.0 \pm 1.4) \times 10^2$, where this value is consistent with the corresponding results found from the figures in the literature \cite{Smith:2009jr} and the error bar is also equal to the Fisher matrix value $1/\sqrt{f_{\rm sky} {\cal N}}$. This is another validation check of our data treatments. 

In the PMF case, our final results from the raw and foreground-clean maps are, respectively, $A = -1.8 \pm 1.4$ and $-1.5 \pm 1.4$ (68\% C.L.), indicating consistency with Gaussianity at $2 \sigma$ regardless of the presence of foregrounds. Taking into account the foreground-cleaned result and a theoretical prior, $A \geq 0$, we find new upper limit on the PMF strength, namely, $B_{\rm 1~Mpc} < 3.1$~nG at $95\%$ C.L. As expected, this is tighter than the passive scalar-mode constraint from {\it Planck} mentioned in Sec.~\ref{sec:theory} ($A = -1.9 \pm 13.9$), owing to considering the tensor-mode contribution. 

%%%%%%%%%%%%%%%%%%%%%%%%%%
\section{Summary and discussion}

The origin of the observed large-scale magnetic fields is one of the most important and interesting issues in probe of the early Universe, and some researchers seek answers in the inflationary paradigm. 
In this paper, we have discussed an observational constraint on the seed magnetic field stretched by the inflationary expansion from the analysis of non-Gaussianities in CMB anisotropies. Signal of the Gaussian field strength of the PMF becomes largest on large scales via the enhancement of non-Gaussian tensor perturbations through the ISW effect. 

We have analyzed the WMAP 7-year temperature maps and confirmed no evidence of squeezed-type tensor non-Gaussianity due to PMFs. Our constraint on the amplitude of the tensor non-Gaussianity leads to an upper bound on the PMF strength as $B_{\rm 1~Mpc} < 3.1 ~{\rm nG}$ (95\% C.L.). This result is not sensitive to the foregrounds. This value may be improved by considering impacts of polarizations \cite{Shiraishi:2013vha}.

Aside from the issue on PMFs, this is a first challenge to constrain a primordial tensor non-Gaussianity from the CMB bispectrum. The tensor CMB bispectrum has spectral shapes quite distinct from the scalar one and leads to nontrivial constraints that have never seen in the scalar case. Unfortunately, the tensor bispectrum has a tangled multipole dependence, and the bispectrum estimator cannot be efficiently factorized as the KSW approach. In the present paper, we have performed ${\cal O}(\ell_{\rm max}^5)$ huge amount of summations to compute the estimator in the brute-force way by focusing solely on large-scale signals up to $\ell_{\rm max} = 100$. In other words, the brute-force method is not applicable to the data analysis with higher resolution, but it will be possible to access such small scales by means of a model-independent factorizable estimator~\cite{Fergusson:2009nv, Fergusson:2010dm, Liguori:2010hx, Fergusson:2011sa, Shiraishi:2014roa, Fergusson:2014gea, Shiraishi:2014ila, Liguori:2014}. Probing tensor non-Gaussianity beyond $\ell_{\rm max} = 100$ remains as a next challenging and exciting issue (although, of course, it is naturally expected that the constraints on the magnetic tensor bispectrum do not vary so much since the signal-to-noise ratio is already saturated at $\ell_{\rm max} = 100$). 

%%%%%%%%%%%%%%%%%%%%%%%%%%%%%%%%%
\begin{acknowledgements}
This work was supported in part by a Grant-in-Aid for JSPS Research under Grant Nos. 22-7477, 25-573 (M.S.), 23-5622 (T.S.), the ASI/INAF Agreement I/072/09/0 for the Planck LFI Activity of Phase E2, and Grant-in-Aid for Nagoya University Global COE Program ``Quest for Fundamental Principles in the Universe: from Particles to the Solar System and the Cosmos" from the Ministry of Education, Culture, Sports, Science and Technology of Japan. We also acknowledge the Kobayashi-Maskawa Institute for the Origin of Particles and the Universe and Nagoya University for providing computing resources useful in conducting the research reported in this paper. Some of the results in this paper have been derived using the HEALPix 
\cite{Gorski:2004by} package.
\end{acknowledgements}

%########################################
% Create the reference section using BibTeX
\bibliography{paper}
%\nocite{*}
\end{document}